\documentclass[useAMS]{mn2e}
\usepackage{lscape}
\usepackage{rotating}
\usepackage{amssymb}
\usepackage{float}

\def\msun{\hbox{M$_\odot$}}

\def\frac{\hbox{f$_{\rm mix}$}}

\def\fenrich{\hbox{f$_{\rm enriched}$}}

\def\t4{\hbox{t$_{\rm 4}$}}

\def\cm3{\hbox{cm$^{-3}$}}

\input psfig.sty
\input epsf.sty
\usepackage{graphicx}
\usepackage{epsfig}
\title[GC Mass Loss and Multiple Populations]
{Globular Cluster Mass Loss in the Context of Multiple Populations}
\author[Bastian \& Lardo] {Nate Bastian$^{1}$ and Carmela Lardo$^{1}$\\
$^{1}$ Astrophysics Research Institute, Liverpool John Moores University, 146 Brownlow Hill, Liverpool L3 5RF, UK\\
}
\date{Accepted. Received; in original form}
\pagerange{\pageref{firstpage}--\pageref{lastpage}}
\pubyear{2015}
\begin{document}
\maketitle
\label{firstpage}
\begin{abstract}
Many scenarios for the origin of the chemical anomalies observed in globular clusters (GCs; i.e., multiple populations) require that GCs were much more massive at birth, up to $10-100\times$, than they are presently.  This is invoked in order to have enough material processed through first generation stars in order to form the observed numbers of enriched stars (inferred to be second generation stars in these models).  If such mass loss was due to tidal stripping, gas expulsion, or tidal interaction with the birth environment, there should be clear correlations between the fraction of enriched stars and other cluster properties, whereas the observations show a remarkably uniform enriched fraction of $0.68\pm0.07$ (from 33 observed GCs).  If interpreted in the heavy mass loss paradigm, this means that all GCs lost the same fraction of their initial mass (between $95-98$\%), regardless of their mass, metallicity,  location at birth or subsequent migration, or epoch of formation. This is incompatible with predictions, hence we suggest that GCs were not significantly more massive at birth, and that the fraction of enriched to primordial stars observed in clusters today likely reflects their initial value.  If true, this would rule out self-enrichment through nucleosynthesis as a viable solution to the multiple population phenomenon.
\end{abstract}

\section{Introduction}
\label{sec:intro}

Most scenarios put forward to explain the observed abundance anomalies in globular clusters (GCs) invoke some form of self-enrichment, where stars produce ``enriched material" through nucleosynthesis (i.e. material that displays the observed abundance variations in light elements) that pollutes other stars within the same cluster.  In the popular ``AGB scenario" (e.g., D'Ercole et al.~2008), AGB stars, formed in a first generation, shed enriched material which, when mixed with primordial material, form a second generation (or more) of stars that show the observed chemical anomalies.  Due to AGB star lifetimes, this is expected to operate over $30-200$~Myr timescales.  The ``Fast Rotating Massive Star" scenario (FRMS - e.g. Decressin et al.~2007; Krause et al.~2013) also invokes multiple star-forming epochs, although the timescales are significantly shorter. This scenario uses the ejecta of rapidly rotating massive stars, when mixed with primordial material left after the formation of the first generation, to form subsequent generations. The timescales associated with the FRMS scenario are $<10$~Myr.

In these scenarios, stars with primordial abundances (i.e. similar to that shown in halo field stars) are interpreted to be from the first generation (FG) while stars that are enriched in some light elements (i.e., Na, Al, He) and depleted in others (C, O) are interpreted to be from a second generation (SG).  We note that the observations only show whether a star is chemically depleted/enriched or not, whereas the labels of first or second generation are interpretations, as there is no direct evidence for the majority of GCs of multiple star-forming events (see the discussion in Chantereau et al.~2015).

One problem that the above scenarios have is that observations show that there are approximately equal numbers of primordial (assumed to be FG in the above scenarios) and enriched (assumed to be SG stars) stars within GCs (e.g., Carretta et al.~2009).  Since the polluting stars only make up a very small fraction of the total initial mass of the cluster ($3-8\%$ - e.g., de Mink et al.~2009) the current observed primordial stars (and their associated higher mass stars that have already burned out) are not enough to have produced the required amount of enriched material.  This is often referred to as the ``mass budget problem".  

A commonly invoked solution to this fundamental problem of self-enrichment scenarios is to have the clusters lose large fractions ($>90$\%) of their FG stars due to tidal field of the Galaxy (e.g., D'Ercole et al.~2008; Conroy 2012) or gas expulsion after the SG has formed (after $\sim10$~Myr for the FRMS scenario or $\sim100$~Myr for the AGB scenario).  This itself would need some fine tuning to result in all clusters having similar enriched stellar fractions, \fenrich, from an initial \fenrich$\lesssim0.05$ to the current \fenrich$=0.5-0.8$, implying that every cluster lost $95-99$\% of their initial FG stars.  While finely tuned, this is, at least in principle, feasible.  



Observations of the ratio of field and globular cluster stars in dwarf galaxies, however, have shown that GCs in those systems could not have been more than a factor of $5-10$ (and potentially substantially less) more massive at birth than they currently are (Larsen et al.~2012; 2014a), calling into question the proposed scenarios.  Schiavon et al.~(in prep.) have found, based on the recent {\em APOGEE} survey, that GCs in the bulge of the Galaxy could only have had initially $\sim8$ times more primordial stars than enriched stars (and likely much less), again in contradiction with the model requirements.   The Larsen et al. and Schiavon et al. constraints are quite complimentary, as they sample very different environments (dwarf galaxies and the bulge of the Milky Way, as well as different metallicity regimes, [Fe/H]$<-2$ and $-2 < [Fe/H] < -1$, respectively).

There are other independent tests that can be carried out on the GC population of the Milky Way in order to test whether the requirements of the self-enrichment models can be satisfied.  One way is to look for the effects of tidal stripping on the multiple populations within current GCs.

If mass loss is due to tidal stripping\footnote{We use the term ``tidal stripping" in a general way to include the loss of GC stars due to two-body relaxation in a tidal field (where stars escape more easily due to the tidal boundary) as well as due to the expansion of the cluster over the tidal boundary.} from the host galaxies, the dissolution time of a cluster is linearly proportional to the Galactocentric distance (assuming a flat rotation curve - e.g., Baumgardt \& Makino~2003; Kruijssen \& Mieske~2009; Lamers et al.~2010).  Hence, for a given mass, clusters in the inner regions of the Milky Way should lose their stars more rapidly than clusters in the outer regions.  Because the scenarios invoke the preferential loss of FG stars, \fenrich\ would increase as the cluster dissolves.  The result should be that clusters in the inner parts of galaxies should have a higher fraction of SG stars, while GCs in the outer parts of the Galaxy should be dominated by their FG stars.  The same applies to GCs formed within dwarf galaxies, as their tidal fields are substantially weaker than the inner Galaxy.

This can be tested by looking for a trend of \fenrich\ with Galactocentric distance.  Di Criscienzo et al.~(2011) studied the Galactic cluster, NGC~2419, a massive ($\sim10^6$~\msun) GC that is one of the most distant ($\sim90$~kpc) of the Milky Way population.  In the strong cluster dissolution paradigm, invoked in the self-enrichment scenarios discussed above, due to its high mass and large Galactocentric distance, this cluster is not expected to have lost many FG stars, so its current ratio should be close to the initial value.  Contrary to expectations,  Di Criscienzo et al.~(2011) found that enriched stars make up $>30$\% of the population, which has been spectroscopically confirmed (Mucciarelli et al.~2012; Cohen \& Kirby~2012).  The authors recognised this discrepancy, and as a potential solution, suggested that  the orbit of NGC~2419 was extremely elliptical, with a peri-centre passage of only $\sim11$~kpc.  However, due to its high mass even at this distance, NGC~2419 would not be expected to lose a large fraction of its stars (e.g., Baumgardt \& Makino 2003).  Hence the high enriched fraction of stars within this cluster remains unexplained.

We can also search for relations between \fenrich\ and metallicity, as it is low-metallicity clusters (in the outer Milky Way) that are expected to be accreted from dwarf galaxies.  This will test whether the birth environment of GCs is responsible for the required mass-loss to make self-enrichment scenarios viable, although we note that current estimates do not find the heavy mass loss required for the majority of GCs (e.g., Kruijssen 2014; 2015).

While any individual GC may have an anomalous orbit (i.e. highly elliptical), it would require extreme fine tuning if other distant GCs were found that had significant fractions of enriched stars.  This would require the GC population has a whole to  have a highly radially anisotropic velocity distribution, which is not consistent with observations (see the discussion in Vesperini et al.~2003).  In the present paper we use data from the literature on the fraction of enriched stars within particular GCs based on large spectroscopic datasets, and supplement this with studies that use HST photometry in the appropriate filters where the different populations can be distinguished.  We then search for the expected radial trend, cluster mass relation, or metallicity influence, which are required in self-enrichment scenarios. The expectation is that outer halo GCs, higher mass GCs, or low metallicty GCs should have a drastically reduced fraction of enriched stars (reflecting their initial fractions).

Finally, we can search for a relation between \fenrich\ and cluster mass.  In the one-dimensional hydrodynamical simulations performed by D'Ercole et al.~(2008) and Vesperini et al.~(2010), the initial fraction of \fenrich\ is a strong function of mass, with lower mass ($10^5$~\msun) clusters initially having \fenrich$=0.01$ and higher mass clusters  ($\sim10^6$~\msun) initially having \fenrich$ > 0.07-0.13$, depending on the initial radius.  Hence, we may expect that some remnant of this relation has been maintained during the subsequent evolution of the GCs. In the present work we adopt an initial value of $f_{\rm enriched}^{\rm initial}=0.05$ from the models of e.g., Vesperini et al.~(2010).

This paper is structured as follows.  In \S~\ref{sec:observations} we introduce the datasets used in the present work while in \S~\ref{sec:disruption} we investigate the expected relations between \fenrich\ and other cluster properties if the clusters have undergone large amounts of mass loss during their evolution.  In \S~\ref{sec:interpretation} we discuss our results in light of the self-enrichment scenarios as well as how our results relate to others in the literature.  Finally, in \S~\ref{sec:conclusions} we present our conclusions.

\section{Observations}
\label{sec:observations}

The primary dataset used in the present work is taken from Carretta et al.~(2009; 2010a).  The authors used a large spectroscopic database of red giant branch stars (RGBs) in a sample of 19 Galactic GCs, and determined the fraction of primordial (P), intermediate (I) and extreme (E) stars., where I and E stars show enrichment patterns not observed in halo field stars.  For the present work, we sum the fractions of the I and E populations, as we are only concerned about the fraction of enriched stars, regardless of the level of enrichment that they display.

To this dataset we include results from a number of recent photometric and spectroscopic samples.  We add the following spectroscopic datasets: NGC~362 (Carretta et al.~2013); NGC~1851 (Carretta et al.~2011); NGC~4833  (Carretta et al.~2014a);  NGC~5286 (Marino et al.~2015); NGC~6093 (Carretta et al.~2015), NGC~6121 (Carretta et al.~2010a); NGC~6752 (Carretta et al.~2012); and NGC 6864 (Kacharov et al.~2013).  We have also included the spectroscopic results of NGC~5272 and NGC~6205 (Sneden et al.~2004; Cohen \& Melendez~2005) which have been analysed in terms of their P, I and E assignment by Carretta et al.~(2009).  

We include the spectroscopic results on the distant and massive GC, NGC~2419, from Cohen \& Kirby~(2012).  If we use [Mg/Fe] to split enriched vs. primordial stars we obtain $\fenrich=0.61$.  If, instead, we use [Na/Fe] as the discriminator we obtain $\fenrich=0.84$.  For the present work we simply take the mean of these two values and adopt $\fenrich=0.73$. We note that in general [Mg/Fe] is not a good tracer of chemical anomalies, as [Mg/Fe] does not vary in many clusters (c.f. Bastian et al.~2013b) that are observed to host multiple populations, hence [Na/Fe] is likely more representative for the cluster.  This is consistent with the lower limits inferred spectroscopically by Mucciarelli et al.~(2012) and photometrically by di Criscienzo et al.~(2011).  These are lower limits, as the two studies were most sensitive to extremely enriched (either depleted in Mg or enriched in He) stars, and likely missed the intermediate population stars.  

Similarly, the results of NGC~6266 (M62) from Yong et al.~(2014) are included, although this point has relatively large errors as it is based on only 7 stars (4 enriched, 3 primordial, resulting in $\fenrich=0.57$).  Just taking Poissonian errors leads to an uncertainty of $\sim0.3$. This result agrees with the photometric study of Milone (2015), who found a lower limit of \fenrich$>0.2$.  This result was a lower limit as the author detected two clearly distinguishable sequences along the main sequence of the cluster, however, the main sequence associated with the ``primordial" population was broader than the photometric errors (consistent with a spread in helium).  Hence, the stars associated with this main sequence likely represent a mixture of primordial and intermediate enriched stars.  Additionally, the photometry presented in Milone (2015) was more sensitive to He spreads than spreads in light element abundances and some clusters with relatively large Na and O spreads display small He spreads (e.g., Bastian et al.~2015).

The use of high precision photometry of GC stars with specific filters that are sensitive to light element abundance variations has the potential to significantly refine the estimates of \fenrich\ based on spectroscopic measurements (e.g., Milone et al.~2015).

We note that we have not homogenised the data (i.e. applied consistent NLTE corrections to the Na abundances), although the majority of the datasets come from the same team, which use consistent techniques.

We also use the results based on HST photometry of NGC~6362 (Dalessandro et al.~2014) and  NGC~7089 (Milone et al.~2015)  

Finally, we include the estimate of D'Antona \& Caloi~(2008) for NGC 6229.  This result is based on the Horizontal Branch morphology from HST photometry.  Since it relies on detailed stellar evolutionary modelling, instead of coming directly from the observations, we consider this to be a relatively uncertain data point.  However, it follows the overall trends for the other clusters in our sample.

In several cases, photometric and spectroscopic datasets exist for the same cluster, and estimates of \fenrich\ have been made using both datasets.  Overall, the agreement between the techniques is good, i.e., generally within the respective error bars (e.g., NGC~6121 - Carretta et al.~2010a and Milone et al.~2014 or NGC~2808 - Carretta et al.~2010a and Piotto et al.~2007).  However, we note that as photometric accuracy improves along with imaging in filters that are good tracers of chemical anomalies, the picture is becoming a bit more complicated.  For example, the red main-sequence identified in NGC~2808 by Piotto et al.~(2007) which is thought to correspond to the ``primordial" stars identified spectroscopically, actually consists of three sub-populations (Milone et al.~2015).  Hence, it is unclear which of the three sub-populations of the ``red-MS" actually corresponds to the primoridal population.  This has also been noted a study of NGC~6752 by Milone et al.~(2013), who found three populations based on photometry, which did not fit neatly into the P, I and E spectroscopic framework of (Carretta et al.~2009a).  However, even here, there was a reasonable agreement in distinguishing between P and I(+E) stars using photometric and spectroscopic methods.  Future studies, using a homogenised analysis of a full sample of GC with the appropriate filter combination (e.g., with the {\em HST UV Legacy Survey of GCs} - Piotto et al.~2015) will compliment and extend the current study.

Based on the comparison between spectroscopic and photometric divisions of the sub-populations, some of the estimated \fenrich\ from spectroscopy may be lower limits.


We have intentionally excluded GCs that are commonly referred to as members of the Sagittarius Dwarf Galaxy as their current position within the Galaxy is not representative of the tidal field experienced by the clusters over a majority of their lifetimes.  We note that this removes M54 and NGC~4590; both which show large enriched populations (Carretta et al.~2010b), but also the low mass clusters Terzan 7, 8 and Palomar 12 which hosts small or non-existent enriched populations according to spectroscopic studies (e.g., Carretta et al.~2014b).  We note that Terzan 7 and Palomar 12 are much younger than typical GCs ($\lesssim9$~Gyr - Leaman et al.~2013) hence age may play a role in whether chemical anomalies are present in clusters, although Terzan 8 appears to be similar in age to the bulk of Galactic GCs (Mar{\'{\i}}n-Franch et al.~2009).  Future imaging of these clusters should provide a better estimate of \fenrich\ for these clusters.

Additionally, we have excluded GCs with evidence of significant spreads in [Fe/H], such as $\omega$-cen, M54, M22 and M19.  This is because clusters with an intrinsic iron spread also display an s-process element bimodality, with the s-rich population also enriched in iron.
Each s-process groups has its own C-N and Na-O anti-correlation (s-rich stars are also on average richer in C, N, and Na with respect to s-poor stars).  In theses cases it is not trivial to distinguish between P, I, and E stars, since the two Na-O anti-correlations overlap (e.g., Marino et al.~2011). 

This results in a sample of 33 GCs with full information available. 

\begin{figure*}
\centering
\includegraphics[width=17cm, angle=0]{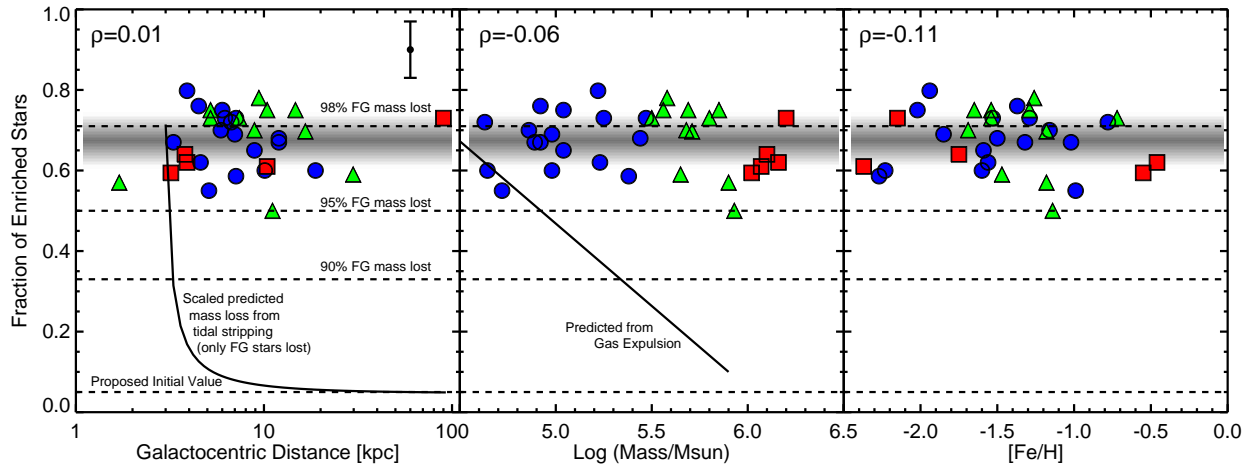}

\caption{The fraction of enriched stars as a function Galactocentric distance (left panel), mass (centre panel), and [Fe/H] (right panel).  Blue (circles), green (triangles), and red (squares) symbols represent GCs with current masses $\lesssim3 \times10^5$\msun,  $3-10 \times 10^5$\msun, and $>10^6$\msun, respectively.  The horizontal dashed lines represent the initial value in the AGB or FRMS scenario, where the SG stars only make up a small fraction of the total mass of the cluster, and subsequent \fenrich, if 90, 95 or 98\% of the FG mass has been removed (from bottom to top, respectively).   The solid line in the left panel shows the expected relation if strong mass loss due to tidal stripping acted on the clusters, preferentially removing only FG stars. This has been scaled to match the observations at $3$~kpc, whereas the actual cluster mass loss is expected to be much weaker. The same relation is expected if stellar evolutionary driven expansion is the cause of the mass-loss through mass segregation.   The solid line in the centre panel is the prediction of Khalaj \& Baumgardt~(2015) if the mass loss was due to gas expulsion.   A representative uncertainty in \fenrich\ is shown in the left panel.  The grey shading denotes the mean and standard deviation of the observed GCs. The Spearman Rank correlation coefficients are given in each panel, no significant correlation between \fenrich\ and Galactocentric distance, mass or metalliicty is found.}
\label{fig:data}
\end{figure*}

We show the resulting trend between the fraction of enriched stars, \fenrich, and Galactocentric distance, GC mass and metallicity  in Fig.~\ref{fig:data}.  Cluster masses were taken from Kruijssen \& Mieske~(2009) or Gnedin \& Ostriker~(1997), while Galactocentric distances and [Fe/H] were taken from Harris~(1996, 2010).  Throughout this work the fractions referred to are that of the long lived low mass stars, in order to be able to compare with observations directly. 

The data show a remarkably constant enriched fraction, with little scatter, $\fenrich=0.68\pm0.07$ (the mean and standard deviation of the sample).  This finding is similar to that found in Carretta et al.~(2009 - their Fig.~10), but now with a larger GC sample.  Carretta et al.~(2009) have estimated the expected scatter, based on the observed numbers of stars and errors associated with assigning stars to the P, I or E sub-populations, of $\sigma(\fenrich)\simeq0.07$ for their sample (making up 19 out of 33 GCs in our sample).  Hence, the observed scatter can be nearly entirely explained within the expected error budget, leaving little room for correlations with other variables.  Least squares fit show that the \fenrich\ is consistent with no relation with Galactocentric radius, mass or metallicity of the clusters.

{\em From the data directly we note that the similar value of  \fenrich\ observed in all clusters argues against any environmentally dependent dynamical process to significantly alter the initial value, as a much larger scatter would be expected.  If interpreted as being due to mass loss, the data suggest that {\em all} clusters lost nearly identical fractions of their initial mass (between $95-99$\%) regardless of their mass, metallicity, or location at birth or subsequent migration.}

\section{Expectations of Cluster Dissolution}
 \label{sec:disruption}
 
 \subsection{Evolution of Clusters in a Tidal Field}
 
 The dissolution of globular clusters due to the combined effects of stellar evolution, two body relaxation within a tidal field and external shocks (such as disk/bulge shocks or close GMC passages) has been extensively studied, both numerically (e.g., Baumgardt \& Makino 2003) and analytically (e.g., Lamers et al.~2010 and references therein).  A basic result of these investigations is that for a given mass and density, clusters in stronger tidal fields dissolve more rapidly than clusters in weaker tidal fields.
 
 The mass budget problem within self-enrichment scenarios has been recognised from an early stage, hence models, such as D'Ercole et al.~(2008), have long invoked cluster dissolution as a potential solution.  These models have maximised the role of dissolution by adopting extreme parameters.  Namely, they have assumed strong tidal fields (e.g., at $\sim4$~kpc from the Galactic centre), very diffuse clusters (half mass radii of $\sim25$~pc - in contrast with young massive clusters and the present properties of GCs - e.g., Portegies Zwart et al.~2010), that the clusters are tidally limited from birth (so that any expansion leads to FG star mass loss), that the clusters were initially mass segregated (to maximise cluster expansion) and that the SG stars were much more concentrated initially than the FG stars (see Khalaj \& Baumgardt~2015 for a further discussion of these points).  If these extreme assumptions are relaxed, the models are not able to explain the observed ratios of enriched and primordial stars (D'Ercole et al.~2008).  
 
Baumgardt \& Makino (2003) have carried out a large suite of N-body simulations, which include stellar evolution (and the resultant stellar evolution driven expansion), of clusters of various masses and at different Galacocentric distances. The authors find that the dissolution time of a cluster is directly proportional to the Galactocentric radius ($t_{\rm dis} \propto R_{\rm GC}$), for a given mass and orbital eccentricity.  This result is also expected from analytical theory of cluster dissolution (e.g., Kruijssen \& Mieske~2009; Lamers et al.~2010, and references therein).  They also found that higher mass clusters have longer dissolution timescales, with $t_{\rm dis} \propto M^{0.6}$.
 
 Since FG stars are generally thought to form in a more extended distribution than SG stars (e.g., Lardo et al.~2011; although see Larsen et al.~2015 for a counter-example), the loss of stars due to cluster dissolution will preferentially remove FG stars.  Therefore, clusters that have undergone more mass loss (i.e., the loss of stars) should have a higher fraction of enriched stars, \fenrich.  Since the dissolution timescale (which is proportional to the rate of mass loss - e.g., Lamers et al.~2010) is a linear function of Galactocentric distance, this naturally suggests that \fenrich\ should be a linear function of Galactocentric distance.
 
Most of the GCs that have had their abundances analysed to date are relatively high mass clusters ($>2 \times 10^5$~\msun), where the tidal stripping is not expected to remove large fractions of the cluster mass, even at relatively small Galactocentric distances ($\sim5-10$~kpc - e.g., Kruijssen~2014).  Hence, most of the GCs studied spectroscopically to date have dissolutions timescales well in excess of a Hubble time (e.g., Baumgardt \& Makino~2003). We note that in the D'Ercole et al.~(2008) model, much of the mass loss is due to stellar evolutionary expansion and not two-body relaxation.  This process (stellar evolutionary driven expansion) is included in the Baumgardt \& Makino~(2003) simulations, but is expected to be weaker due to the less extreme initial conditions (half-light radii, mass segregation) assumed by Baumgardt \& Makino~(2003).   

Even in the case of mass-loss driven by stellar evolutionary expansion, the amount of mass lost is expected to be inversely proportional to GC mass and Galactocentric radius, although it may be somewhat different in magnitude relative to two-body relaxation driven tidal stripping.  The effects of primordial mass segregation, and its role in cluster dissolution through stellar evolutionary induced cluster expansion, has been studied numerically by Haghi et al.~(2014).  The authors carried out a suite of N-body simulations of clusters in a Milky Way type potential, for clusters with and without mass segregation.   As expected, they find that mass-segregated clusters dissolve more rapidly (by a factor of $\sim4$), due to stellar evolutionary driven expansion.  However, they also find that the dissolution timescale for mass segregated and non-mass segregated clusters depends on Galactocentric radius in the same way, with $T_{\rm dis} \propto R_{GC}^{\alpha}$, and $\alpha = 1.1-1.3$.  This is similar to that found by Baumgardt \& Makino~(2003).  Hence, whether cluster mass-loss is driven mainly by two-body relaxation or stellar evolutionary driven expansion in a tidal field, the fraction of mass loss (and subsequently \fenrich\ for the models discussed here) should be a linear function of Galactocentric distance.
 
 Although studies have shown that GCs are not expected to have lost large fractions of their initial mass due to tidal stripping (e.g., Baumgardt \& Makino, Kruijssen \& Mieske~2009; Lamers et al.~2010; Kruijssen~2014), we are not interested in testing the absolute timescales for cluster dissolution in the present work.  Rather we will focus on its expected dependence with Galactocentric distance.  As such, we assume that cluster dissolution operates in such a way as to remove 98\% of the FG stars at a radius of $3$~kpc, and hence match the mean of the observations at that radius.  In Fig.~\ref{fig:data}, left panel, we show the expected decline in \fenrich\ as a function of Galactocentric distance, where the fraction of FG stars lost depends linearly on distance.  This technique also removes much of the dependence of GC dissolution on cluster mass, unless there is a strong relation between mass and Galactocentric data.  In Fig.~\ref{fig:data} the GCs are colour-coded in three mass bins (see caption for details), and no strong dependence on cluster mass is apparent.

Such tidal stripping should also depend on the mass of the GC, with higher mass GCs undergoing less mass loss ($t_{\rm dis} \propto M^{0.7}$ - e.g., Baumgardt \& Makino~2003; Lamers et al.~2010).  Hence, we would expect that higher mass GCs should have low \fenrich\ values.  In the centre panel of Fig.~\ref{fig:data} we show the observed relation between \fenrich and cluster mass.  No correlation is identified, in contrast with model predictions.
 
Additionally, we note that if clusters were substantially larger in the past, the amount of mass lost through two-body relaxation and tidal stripping would be much less, and even tidal shocks (bulge, disk or GMCs) would not be expected to remove much mass.
 
We have explicitly assumed that only FG stars are lost (which maximises the efficiency of tidal stripping to bring the initial \fenrich\ up to the current values), although even in the outer regions of GCs today, enriched stars are only slightly more centrally concentrated than stars with primordial abundances (e.g., Lardo et al.~2011; Vanderbeke et al.~2015), hence in practice we would expect some fraction of SG stars to be lost in addition to FG stars, in the self-enrichment plus mass-loss models.
 
In the above estimates we have assumed (as in previous works, e.g., D'Ercole et al.~2008) that a cluster's current orbit is representative of the average tidal field experienced by the GC over its lifetime.  However, this may not be valid as many, especially the low metallicity, GCs are thought to have been accreted from dwarf galaxies (e.g., Brodie \& Strader~2006).  As such, we may expect a relation between the metallicity of a GC (as a proxy of whether it formed in-situ or was later accreted from a dwarf galaxy) and \fenrich.  The tidal field of a dwarf galaxy is substantially weaker than that of the inner Milky Way, so whether a GC spent most of its life orbiting at large Galactocentric radii, or was accreted from a low-mass dwarf (which in turn corresponds, on average, to the GCs at large Galactocentric radii), we would expect these clusters to have lost less of their FG stars than inner Galaxy GCs.  

In the right panel of Fig.~\ref{fig:data} we show \fenrich\ against [Fe/H] for our cluster sample.  As was found for Galactocentric distance and GC mass, there is no clear relation between the fraction of enriched stars and metallicity, so the birth location or subsequent migration has not influenced \fenrich.

{\em Hence, we conclude that if mass-loss is driving the observed value of \fenrich, then the mass loss is independent of Galactocentric radius, metallicity, current mass and birth environment (and subsequent migration), in contradiction with all mass loss mechanisms proposed so far.} 
 
 \subsection{Gas Expulsion}
 \label{sec:gas_expulsion}
 
 Khalaj \& Baumgardt~(2015) have investigated, through N-body simulations, whether the observed high fractions of enriched stars could be due to mass loss induced by gas expulsion (following on the work of Decressin et al.~2010).  If the SG stars form from a gas reservoir that contains a substantial fraction of the total mass of the cluster (i.e. similar to the total mass of FG stars), and if this gas is removed rapidly from the cluster, the resultant expansion of the cluster can result in the loss of large fractions of FG stars (we again are assuming to be less centrally concentrated than SG stars).
 
 Note that this is not the gas expulsion after the formation of the FG, as that would remove the stars too early, and not allow them to contribute to the material needed to form the SG stars.
 
 Khalaj \& Baumgardt~(2015) find that in order to reproduce the observed \fenrich, a considerable amount of fine tuning is necessary.  Only a small part of parameter space was able to lead to clusters that 1) survived and 2) had \fenrich between 0.5 and 0.8.  However, for the surviving clusters, their models would predict a strong correlation between the mass of the cluster and \fenrich.  This is due to the fact that in order to have large \fenrich\ ratios, GCs need to lose a large fraction of the total mass, resulting in lower mass clusters on average.  
  
 Their prediction (approximate linear fit to the models presented in their Fig.~6) is shown compared to the data in the middle panel of Fig.~\ref{fig:data}.  We see that the expected relation between cluster mass and \fenrich\ is not followed by the observations.  Hence, we conclude, like Khalaj \& Baumgardt~(2015) that gas expulsion following the formation of the SG is unlikely to be a viable mechanism to remove large fractions of FG stars.
 
A similar argument could be made for the ``cruel cradle" mass loss mechanism (c.f., Kruijssen~2014), where the gas rich environment (made up of GMCs) where clusters form, also rapidly destroys clusters (due to GMC interactions).  This mechanism is expected to operate too quickly for the AGB scenario, as it is most pronounced in the first 10-20~Myr of a cluster's life.  Regardless, clusters with high \fenrich\ would be expected to be lower-mass, on average, as they would be the clusters that have undergone the most mass loss.  Hence, if this mechanism was responsible, we would expect strong trends of \fenrich\ with cluster mass, and potentially metallicity as this may reflect the birth environment, neither of which are seen in the observations.

\section{Discussion}
 \label{sec:interpretation}
 
 \subsection{Constraining the Mass Loss from GCs}
 
 The observations show that \fenrich\  is remarkably constant ($0.68\pm0.07$).  If the current value of \fenrich\ is interpreted as being due to mass loss, the data suggest that {\em all} clusters lost nearly identical fractions of their initial mass (between $95-99$\%) regardless of their mass, metallicity, or location at birth or subsequent migration, requiring a high level of fine-tuning.

We have searched for dependencies of \fenrich\ on other cluster properties, which are expected to be present if clusters have lost large fractions of their initial masses.  We find that it is independent of Galactocentric distance, and does not follow the expected trend of decreasing \fenrich\ with Galactocentric radius that would indicate that a significant amount of mass loss has occurred due to tidal stripping (either through two-body relaxation or cluster expansion driven by stellar evolutionary mass loss in a mass segregated cluster).  Similar results are found for the case of gas expultion, as \fenrich\ does not follow the predicted trend with GC mass.  We also do not find any significant trend between \fenrich\ and cluster metallicity, [Fe/H].  Our results are in agreement with the previous analyses performed by Carretta et al.~(2010a) and Khalaj \& Baumgardt~(2015).  We also note that GCs 1, 2, 3, and 5 in the Fornax Dwarf Galaxy have similar \fenrich\ values (Larsen et al. ~2014b).
  
The constancy of \fenrich\ as a function of Galactocentric distance and [Fe/H] shows that GCs belonging to the halo and the bulge do not differ in their present enriched fractions.  This is relevant as metal poor halo clusters are often thought to be accreted from dwarf galaxies while metal rich bulge clusters are thought to have formed in-situ (e.g., Brodie \& Strader~2006).  Hence, regardless of the formation epoch or location (and subsequent migration), GCs contain similar fractions of enriched stars.
  
Our results suggest that GCs were not initially significantly more massive than at present, or at least did not lose their stars due to the influence of the tidal field or gas expulsion (or birth location), contrary to the requirements of the self-enrichment scenarios discussed in \S~1. This, in turn, implies that the ``mass budget problem" cannot be solved by increasing the initial mass in clusters.

If the clusters did not undergo a significant amount of preferential mass loss, then the fraction of enriched to primordial stars observed in clusters likely reflects the initial value.  If this is the case, then none of the self-enrichment scenarios can work, without extreme IMF variations, it is not possible to produce the amount of material required to form the observed number of SG stars (even when considering the fraction of primordial material required to be mixed with the polluted ejecta) through standard nucleosynthetic channels.  Additionally, if the current value of \fenrich\ reflects the initial value, the Galactic halo would not be made of significant numbers of lost FG GC stars\footnote{Carretta et al.~(2010a) find that $\sim1.4$\% of the halo may be made up of enriched stars that have escaped GCs.  A similar fraction has been found by Martell et al.~(2011) and Ram{\'{\i}}rez et al.~(2012).}.  In turn, GCs, due to their lower initial masses, would have contributed substantially less to the reionisation of the Universe at high redshift than suggested in the heavy mass-loss paradigm (c.f., Schaerer \& Charbonnel~2011).

\subsection{Other Potential Solutions to the Mass Budget Problem}

Some recent models have attempted to solve the mass budget problem by having young GCs accrete enriched material from their surroundings, i.e., AGB processed material from other stars in the host galaxy not stars within the GC (e.g., Maxwell et al.~2014).  However, this scenario is disfavoured due to the lack of Fe spreads in most GCs observed to date, as the host galaxy ISM is expected to be enriched by both material processed through AGB stars, as well as high mass stars and their associated SNe products.  Even small amounts of material processed in SNe should result in substantial Fe spreads, especially in low-metallicity clusters (e.g., Renzini~2013), so it is unlikely that the accreted material would not introduce (at least small) Fe spreads within the GCs.  Additionally, the ISM will be polluted with the ejecta of lower-mass AGB stars, which do not match the abundance trends observed in GCs in general (D'Ercole et al.~2010).

Alternatively, Charbonnel et al.~(2014) have suggested a solution to the mass budget problem by forming the FG with a stellar IMF devoid of stars below $15-20$~\msun (i.e., the FG only consists of  stars that contribute to the processed material needed to form the SG).  Hence, all stars observed today would be SG stars, even the ones with ``FG-like" abundances.  Since there are no long-lived FG stars, having the FG be 10-100 times more massive than the current mass of the clusters would not contradict the current observations (unless dark remnants of the FG remained which would affect the M/L of the clusters, but these are assumed to be ejected due to SNe kicks).  In this scenario, \fenrich\ would represent the fraction of SG stars born with abundances observably different than the SG stars born primarily from primordial material.  However, in this scenario large He spreads are expected (e.g., Chantereau et al.~2015), contrary to observations (Milone~2015; Bastian et al.~2015) and no young massive clusters with such anomalous stellar IMFs have been observed to date (e.g., Bastian et al.~2010).

We note that resorting to such an anomalously top-heavy stellar IMF is unlikely to help in the AGB scenario.  Unlike in the FRMS scenario, the polluting stars do not end their lives as SNe, and hence will not receive strong kicks at the end of their lives that could potentially remove them from the cluster. Such remnants would leave a strong dynamical signature within clusters as well an anomalous population of white dwarfs, which would be inconsistent with observations of white dwarfs in some GCs (Richer et al.~2008).  In the standard AGB scenario, the IMF is already assumed to be highly anomalous, as the SG only consists of stars that would remain alive until the present day, i.e. only stars with masses below 0.8~\msun.  Allowing for a normal IMF for the SG stars would result in an additional factor of two in the mass-budget problem (see Cabrera-Ziri et al.~2015 for a further discussion of this point and other underlying assumptions made to minimise the mass-budget problem in the AGB scenario).

\subsection{Constraints from Other Sources}

Our results show that the observations are inconsistent with the proposed self-enrichment models, as GCs are unlikely to have lost substantial fractions of their initial masses.  Similar conclusions have also been reached independently from a variety of sources.  Along similar lines as the present work, Larsen et al.~(2012; 2014a) have shown that GCs in a sample of three dwarf galaxies could not have been more than $5-10$ times more massive at birth than their present mass, in tension with the requirements of self-enrichment models.  Similar results have also been found for the Galactic Bulge (Schiavon et al.~in prep.).

The above results also place constraints on the size of the initial GC population, namely whether it was much larger than observed today.  Since the observed field stars in dwarf galaxies and the Galactic bulge are already not numerous enough to produce enough enriched material to form the observed enriched (second generation) stars, there is little room for a large initial GC population where the majority of GCs are completely disrupted.  Hence the current GCs are not just the surviving minority (due to some special initial conditions) of an initially much larger GC population.

The main reason why nuclear burning has been invoked to explain the origin of multiple populations, is that many of the elements observed or inferred to vary from star-to-star within GCs are associated with hot hydrogen burning.  However, in such processing, the abundance of He (the main product of hydrogen burning) is expected to be directly linked with changes in other elements, such as Na, O, C, and N.  Recently, Bastian et al.~(2015) have shown that the observed extent of the Na-O correlations in clusters is not directly linked to the He spreads within the clusters, in contradiction to basic nucleosynthesis.  Additionally, they showed that if an enriching source can explain clusters like NGC~2808, it will necessarily fail to reproduce more typical GCs like NGC~104 (47 Tuc), and vice-versa, unless there is a strong stochastic element to the model.  This is fundamentally at odds with all the proposed enrichment sources (AGBs, FRMSs, very massive stars, or interacting binaries) and appears to rule out all nucleosynthetic sources as the origin of the abundance anomalies.  

All self-enrichment models proposed to date predict that the enriched stars should be more centrally concentrated than the primordial stars (modulo dynamical evolution).   However, the observed radial profiles of some GCs (e.g., M15 and NGC~2808) are inconsistent with predictions as the primordial stars are more centrally concentrated than the enriched stars (M15 - Larsen et al.~2015 ) or that all populations have the same distribution (NGC~2808 - 
Iannicola et al.~2009; Dalessandro et al.~2011). These discrepancies remain even after considering the effects of dynamical evolution. 

Since the proposed scenarios do not invoke any special physics (i.e. conditions in the early universe) the processes should be observable in massive clusters forming today.  However, to date, no ongoing secondary episodes of star-formation within these clusters has been found  (Bastian et al.~2013a; Cabrera-Ziri et al.~2014) nor have the gas reservoirs required to form the secondary population been found (Bastian \& Strader~2014; Cabrera-Ziri et al.~2015; Longmore~2015).  

Additionally, observations have found unexpected correlations between [N/Fe] and GC mass in the M31 GC population (Schiavon et al.~2013) as well as a strong trend of He spread with GC mass in the Galaxy (Milone~2015).  These are not predicted in any of the self-enrichment models, and can only brought into agreement with models by making the mass-budget problem significantly worse (by additional factors of three or more - see Bastian et al.~2015). Finally, the discovery of the chemical anomalies in bulge stars of the Galaxy (that do not appear to be from dissolved clusters - Schiavon et al.~in prep.) suggests that whatever process that causes the multiple population phenomenon is not limited to GCs, but rather may be a relic of high-redshift star-formation in dense environmnets. 
 
It is still unclear what the controlling factor is in determining whether a cluster hosts chemical anomalies.  The standard idea is that cluster mass is the key parameter, although this was largely based on the idea that GCs were originally much more massive than they are today.  The observations presented here, along with studies of the stellar populations of dwarf galaxies (Larsen et al.~2012; 2014a) show that GCs are unlikely to have been significantly more massive in the past, hence their current masses (and \fenrich) should approximately trace their initial values.   Since young and intermediate age ($<2-3$~Gyr) clusters with masses similar to that of GCs ($\lesssim2 \times10^5$\msun) have not been found to display chemical anomalies (Mucciarelli et al.~2008; 2011; 2014; Davies et al.~2009; Mackey et al.~in prep.), it appears that GC mass is not the key parameter.
  
 \section{Conclusions}
 \label{sec:conclusions}
 
 We have shown that the fraction of enriched stars in Galactic globular clusters is surprisingly uniform, $\fenrich=0.68\pm0.07$, with the scatter being nearly entirely accounted for within the error budget (both in numbers and the associated error of assigning if a star is enriched or not), leaving little room for correlations with other parameters.
 
Scenarios that invoke heavy mass loss (preferentially from the FG) from young globular clusters in order to solve the mass-budget problem, have suggested that the mass-loss is due to tidal stripping, gas expulsion, or the influence of the birth environment.  Such heavy mass loss is expected to result in correlations between the observed fraction of SG-to-FG stars and other cluster properties, such as their mass, metallicity, or Galactocentric distance.  

We have collected data from the literature for 33 GCs and no such correlations are found, suggesting that clusters have not undergone such radical mass-loss during their lives.  The fraction of enriched stars is independent of the GC's metallicity, Galactocentic distance, and mass, from which we infer that it is  also independent of their birth location (in the Galaxy or accreted from dwarfs) or epoch of formation.  

Hence, the observed constancy of \fenrich\ may reflect the initial value, which would exclude self-enrichment through nucleosynthesis as a potential solution to the puzzle of multiple populations, as there would not be enough 'first generation' stars to produce the material require to for the observed 'second generation' stars.
 
Given the results presented here, along with recent results from the literature, it appears that the mechanism responsible for the chemical anomalies observed in GCs is not self-enrichment, at least not through the proposed nucleosynthetic channels.  Therefore alternative theories need to be explored.

\section*{Acknowledgments}

We thank Diederik Kruijssen, Henny Lamers, Michele Bellazzini, Holger Baumgardt, S{\o}ren Larsen, Martin Krause, and Chris Usher for helpful discussions.  We also thank the referee for her/his comments which helped improve the manuscript.  NB is partially funded by a Royal Society University Research Fellowship and a European Research Council Consolidator Grant.

\bsp
\label{lastpage}

\begin{thebibliography}{99}

\bibitem[Bastian et 
al.(2010)]{2010ARA&A..48..339B} Bastian, N., Covey, K.~R., \& Meyer, M.~R.\ 2010, ARA\&A, 48, 339 

\bibitem[Bastian et al.(2013)]{2013MNRAS.436.2852B} Bastian, N., 
Cabrera-Ziri, I., Davies, B., \& Larsen, S.~S.\ 2013a, MNRAS, 436, 2852 

\bibitem[Bastian et al.(2013)]{2013MNRAS.436.2398B} Bastian, N., Lamers, 
H.~J.~G.~L.~M., de Mink, S.~E., et al.\ 2013b, MNRAS, 436, 2398 

\bibitem[Bastian 
\& Strader(2014)]{2014MNRAS.443.3594B} Bastian, N., \& Strader, J.\ 2014, MNRAS, 443, 3594 

\bibitem[Bastian et al.(2015)]{2015MNRAS.449.3333B} Bastian, N., 
Cabrera-Ziri, I., \& Salaris, M.\ 2015, MNRAS, 449, 3333 

\bibitem[Brodie 
\& Strader(2006)]{2006ARA&A..44..193B} Brodie, J.~P., \& Strader, J.\ 2006, ARAA, 44, 193 


\bibitem[Baumgardt 
\& Makino(2003)]{2003MNRAS.340..227B} Baumgardt, H., \& Makino, J.\ 2003, MNRAS, 340, 227 

\bibitem[Cabrera-Ziri et al.(2014)]{2014MNRAS.441.2754C} Cabrera-Ziri, I., 
Bastian, N., Davies, B., et al.\ 2014, MNRAS, 441, 2754 

\bibitem[Cabrera-Ziri et al.(2015)]{2015MNRAS.448.2224C} Cabrera-Ziri, I., 
Bastian, N., Longmore, S.~N., et al.\ 2015, MNRAS, 448, 2224 

\bibitem[Carretta et 
al.(2009)]{2009A&A...505..117C} Carretta, E., Bragaglia, A., Gratton, R.~G., et al.\ 2009, A\&A, 505, 117 

\bibitem[Carretta et 
al.(2010)]{2010A&A...516A..55C} Carretta, E., Bragaglia, A., Gratton, R.~G., et al.\ 2010a, A\&A, 516, A55 

\bibitem[Carretta et 
al.(2010)]{2010A&A...520A..95C} Carretta, E., Bragaglia, A., Gratton, R.~G., et al.\ 2010b, A\&A, 520, A95 


\bibitem[Carretta et 
al.(2011)]{2011A&A...533A..69C} Carretta, E., Lucatello, S., Gratton, R.~G., Bragaglia, A., \& D'Orazi, V.\ 2011, A\&A, 533, A69 

\bibitem[Carretta et al.(2012)]{2012ApJ...750L..14C} Carretta, E., 
Bragaglia, A., Gratton, R.~G., Lucatello, S., 
\& D'Orazi, V.\ 2012, ApJL, 750, L14 

\bibitem[Carretta et 
al.(2013)]{2013A&A...557A.138C} Carretta, E., Bragaglia, A., Gratton, R.~G., et al.\ 2013, A\&A, 557, A138 

\bibitem[Carretta et 
al.(2014)]{2014A&A...564A..60C} Carretta, E., Bragaglia, A., Gratton, R.~G., et al.\ 2014a, A\&A, 564, A60 

\bibitem[Carretta et 
al.(2014)]{2014A&A...561A..87C} Carretta, E., Bragaglia, A., Gratton, R.~G., et al.\ 2014b, A\&A, 561, A87 

\bibitem[Carretta et al.(2015)]{2015arXiv150303074C} Carretta, E., 
Bragaglia, A., Gratton, R.~G., et al.\ 2015, A\&A, in press (arXiv:1503.03074)

\bibitem[Chantereau et al.(2015)]{2015arXiv150401878C} Chantereau, W., 
Charbonnel, C., \& Decressin, T.\ 2015, A\&A in press (arXiv:1504.01878)

\bibitem[Charbonnel et 
al.(2014)]{2014A&A...569L...6C} Charbonnel, C., Chantereau, W., Krause, M., Primas, F., \& Wang, Y.\ 2014, A\&A, 569, L6 

\bibitem[Cohen 
\& Mel{\'e}ndez(2005)]{2005AJ....129..303C} Cohen, J.~G., \& Mel{\'e}ndez, J.\ 2005, AJ, 129, 303 

\bibitem[Cohen 
\& Kirby(2012)]{2012ApJ...760...86C} Cohen, J.~G., \& Kirby, E.~N.\ 2012, ApJ, 760, 86 

\bibitem[Conroy(2012)]{2012ApJ...758...21C} Conroy, C.\ 2012, ApJ, 758, 21 


\bibitem[Dalessandro et al.(2011)]{2011MNRAS.410..694D} Dalessandro, E., 
Salaris, M., Ferraro, F.~R., et al.\ 2011, MNRAS, 410, 694 

\bibitem[Dalessandro et al.(2014)]{2014ApJ...791L...4D} Dalessandro, E., 
Massari, D., Bellazzini, M., et al.\ 2014, ApJL, 791, L4 

\bibitem[D'Antona 
\& Caloi(2008)]{2008MNRAS.390..693D} D'Antona, F., \& Caloi, V.\ 2008, MNRAS, 390, 693 

\bibitem[Davies et al.(2009)]{2009ApJ...696.2014D} Davies, B., Origlia, L., 
Kudritzki, R.-P., et al.\ 2009, ApJ, 696, 2014 

\bibitem[Decressin et 
al.(2007)]{2007A&A...464.1029D} Decressin, T., Meynet, G., Charbonnel, C., Prantzos, N., \& Ekstr{\"o}m, S.\ 2007, A\&A, 464, 1029 

\bibitem[Decressin et 
al.(2010)]{2010A&A...516A..73D} Decressin, T., Baumgardt, H., Charbonnel, C., \& Kroupa, P.\ 2010, A\&A, 516, A73 

\bibitem[D'Ercole et al.(2008)]{2008MNRAS.391..825D} D'Ercole, A., 
Vesperini, E., D'Antona, F., McMillan, S.~L.~W., 
\& Recchi, S.\ 2008, MNRAS, 391, 825 

\bibitem[D'Ercole et al.(2010)]{2010MNRAS.407..854D} D'Ercole, A., 
D'Antona, F., Ventura, P., Vesperini, E., 
\& McMillan, S.~L.~W.\ 2010, MNRAS, 407, 854 

\bibitem[de Mink et 
al.(2009)]{2009A&A...507L...1D} de Mink, S.~E., Pols, O.~R., Langer, N., \& Izzard, R.~G.\ 2009, A\&A, 507, L1 

\bibitem[di Criscienzo et al.(2011)]{2011MNRAS.414.3381D} di Criscienzo, 
M., D'Antona, F., Milone, A.~P., et al.\ 2011, MNRAS, 414, 3381 

\bibitem[Gnedin 
\& Ostriker(1997)]{1997ApJ...474..223G} Gnedin, O.~Y., \& Ostriker, J.~P.\ 1997, ApJ, 474, 223 

\bibitem[Haghi et al.(2014)]{2014MNRAS.444.3699H} Haghi, H., Hoseini-Rad, 
S.~M., Zonoozi, A.~H., K\"upper, A.~H.~W.\ 2014, MNRAS, 444, 3699 


\bibitem[Harris(1996)]{1996AJ....112.1487H} Harris, W.~E.\ 1996, AJ, 112, 
1487 

\bibitem[Harris(2010)]{2010arXiv1012.3224H} Harris, W.~E.\ 2010, 
(arXiv:1012.3224 )

\bibitem[Iannicola et al.(2009)]{2009ApJ...696L.120I} Iannicola, G., 
Monelli, M., Bono, G., et al.\ 2009, ApJL, 696, L120 


\bibitem[Khalaj 
\& Baumgardt(2015)]{2015arXiv150605303K} Khalaj, P., \& Baumgardt, H.\ 2015, MNRAS, in press (arXiv:1506.05303) 

\bibitem[Kacharov et 
al.(2013)]{2013A&A...554A..81K} Kacharov, N., Koch, A., \& McWilliam, A.\ 2013, A\&A, 554, A81 

\bibitem[Krause et 
al.(2013)]{2013A&A...552A.121K} Krause, M., Charbonnel, C., Decressin, T., Meynet, G., \& Prantzos, N.\ 2013, A\&A, 552, A121 

\bibitem[Kruijssen 
\& Mieske(2009)]{2009A&A...500..785K} Kruijssen, J.~M.~D., \& Mieske, S.\ 2009, A\&A, 500, 785 

\bibitem[Kruijssen(2014)]{2014CQGra..31x4006K} Kruijssen, J.~M.~D.\ 2014, 
Classical and Quantum Gravity, 31, 244006 

\bibitem[Kruijssen(2015)]{2014CQGra..31x4006K} Kruijssen, J.~M.~D.\ 2015, MNRAS, submitted

\bibitem[Lamers et al.(2010)]{2010MNRAS.409..305L} Lamers, H.~J.~G.~L.~M., 
Baumgardt, H., \& Gieles, M.\ 2010, MNRAS, 409, 305 

\bibitem[Lardo et 
al.(2011)]{2011A&A...525A.114L} Lardo, C., Bellazzini, M., Pancino, E., et al.\ 2011, A\&A, 525, A114 

\bibitem[Larsen et 
al.(2012)]{2012A&A...544L..14L} Larsen, S.~S., Strader, J., \& Brodie, J.~P.\ 2012, A\&A, 544, L14 

\bibitem[Larsen et 
al.(2014)]{2014A&A...565A..98L} Larsen, S.~S., Brodie, J.~P., Forbes, D.~A., \& Strader, J.\ 2014a, A\&A, 565, A98 
\bibitem[Larsen et al.(2014)]{2014ApJ...797...15L} Larsen, S.~S., Brodie, 
J.~P., Grundahl, F., \& Strader, J.\ 2014b, ApJ, 797, 15 


\bibitem[Larsen et al.(2015)]{2015ApJ...804...71L} Larsen, S.~S., 
Baumgardt, H., Bastian, N., et al.\ 2015, ApJ, 804, 71 

\bibitem[Leaman et al.(2013)]{2013MNRAS.436..122L} Leaman, R., VandenBerg, 
D.~A., \& Mendel, J.~T.\ 2013, MNRAS, 436, 122 

\bibitem[Longmore(2015)]{2015MNRAS.448L..62L} Longmore, S.~N.\ 2015, 
MNRAS, 448, L62 

\bibitem[Mar{\'{\i}}n-Franch et al.(2009)]{2009ApJ...694.1498M} 
Mar{\'{\i}}n-Franch, A., Aparicio, A., Piotto, G., et al.\ 2009, ApJ, 694, 
1498 

\bibitem[Marino et 
al.(2011)]{2011A&A...532A...8M} Marino, A.~F., Sneden, C., Kraft, R.~P., et al.\ 2011, A\&A, 532, A8 

\bibitem[Marino et al.(2015)]{2015MNRAS.450..815M} Marino, A.~F., Milone, 
A.~P., Karakas, A.~I., et al.\ 2015, MNRAS, 450, 815 

\bibitem[Martell et 
al.(2011)]{2011A&A...534A.136M} Martell, S.~L., Smolinski, J.~P., Beers, T.~C., \& Grebel, E.~K.\ 2011, A\&A, 534, A136 


\bibitem[Maxwell et al.(2014)]{2014MNRAS.439.2043M} Maxwell, A.~J., 
Wadsley, J., Couchman, H.~M.~P., \& Sills, A.\ 2014, MNRAS, 439, 2043 


\bibitem[Milone et al.(2013)]{2013ApJ...767..120M} Milone, A.~P., Marino, 
A.~F., Piotto, G., et al.\ 2013, ApJ, 767, 120 

\bibitem[Milone et al.(2014)]{2014MNRAS.439.1588M} Milone, A.~P., Marino, 
A.~F., Bedin, L.~R., et al.\ 2014, MNRAS, 439, 1588 

\bibitem[Milone et al.(2015)]{2015MNRAS.447..927M} Milone, A.~P., Marino, 
A.~F., Piotto, G., et al.\ 2015, MNRAS, 447, 927 

\bibitem[Milone(2015)]{2015MNRAS.446.1672M} Milone, A.~P.\ 2015, MNRAS, 
446, 1672 

\bibitem[Mucciarelli et al.(2008)]{2008AJ....136..375M} Mucciarelli, A., 
Carretta, E., Origlia, L., \& Ferraro, F.~R.\ 2008, AJ, 136, 375 

\bibitem[Mucciarelli et al.(2011)]{2011MNRAS.413..837M} Mucciarelli, A., 
Cristallo, S., Brocato, E., et al.\ 2011, MNRAS, 413, 837 

\bibitem[Mucciarelli et al.(2012)]{2012MNRAS.426.2889M} Mucciarelli, A., 
Bellazzini, M., Ibata, R., et al.\ 2012, MNRAS, 426, 2889 


\bibitem[Mucciarelli et al.(2014)]{2014ApJ...793L...6M} Mucciarelli, A., 
Dalessandro, E., Ferraro, F.~R., Origlia, L., 
\& Lanzoni, B.\ 2014, ApJL, 793, L6 

\bibitem[Piotto et al.(2007)]{2007ApJ...661L..53P} Piotto, G., Bedin, 
L.~R., Anderson, J., et al.\ 2007, ApJL, 661, L53 

\bibitem[Portegies Zwart et 
al.(2010)]{2010ARA&A..48..431P} Portegies Zwart, S.~F., McMillan, S.~L.~W., \& Gieles, M.\ 2010, ARAA, 48, 431 

\bibitem[Ram{\'{\i}}rez et al.(2012)]{2012ApJ...757..164R} Ram{\'{\i}}rez, 
I., Mel{\'e}ndez, J., \& Chanam{\'e}, J.\ 2012, ApJ, 757, 164 

\bibitem[Renzini(2013)]{2013MmSAI..84..162R} Renzini, A.\ 2013, Mem.S.A.It., 
84, 162 

\bibitem[Richer et al.(2008)]{2008AJ....135.2141R} Richer, H.~B., Dotter, 
A., Hurley, J., et al.\ 2008, AJ, 135, 2141 

\bibitem[Schaerer 
\& Charbonnel(2011)]{2011MNRAS.413.2297S} Schaerer, D., \& Charbonnel, C.\ 2011, MNRAS, 413, 2297 

\bibitem[Schiavon et al.(2013)]{2013ApJ...776L...7S} Schiavon, R.~P., 
Caldwell, N., Conroy, C., et al.\ 2013, ApJL, 776, L7 

\bibitem[Schiavon et al.(2014)]{temp} Schiavon, R.~P., et al.\ 2015, ApJ (submitted) 

\bibitem[Sneden et al.(2004)]{2004AJ....127.2162S} Sneden, C., Kraft, 
R.~P., Guhathakurta, P., Peterson, R.~C., 
\& Fulbright, J.~P.\ 2004, AJ, 127, 2162 

\bibitem[Vanderbeke et al.(2015)]{2015arXiv150406509V} Vanderbeke, J., De 
Propris, R., De Rijcke, S., et al.\ 2015, MNRAS, in press (arXiv:1504.06509)

\bibitem[Vesperini et al.(2003)]{2003ApJ...593..760V} Vesperini, E., Zepf, 
S.~E., Kundu, A., \& Ashman, K.~M.\ 2003, ApJ, 593, 760 

\bibitem[Vesperini et al.(2010)]{2010ApJ...718L.112V} Vesperini, E., 
McMillan, S.~L.~W., D'Antona, F., \& D'Ercole, A.\ 2010, ApJL, 718, L112 

\bibitem[Yong et al.(2014)]{2014MNRAS.439.2638Y} Yong, D., Alves Brito, A., 
Da Costa, G.~S., et al.\ 2014, MNRAS, 439, 2638 









\end{thebibliography}
\end{document}